\documentclass[pra,showpacs,floatfix,superscriptaddress,aps,twocolumn]{revtex4}
\usepackage{amsmath}
\usepackage{amssymb}

\usepackage{graphicx}
\usepackage[usenames]{color}

\newcommand{\jpb}{J. Phys. B\ }

\begin{document}

\title{{Dipolar Bose-Einstein condensate for large scattering length}}

\author{Luis E. Young-S\footnote{Email: lyoung@ift.unesp.br}}
\affiliation{Instituto de F\'{\i}sica Te\'orica, UNESP - Universidade Estadual Paulista, 01.140-070 S\~ao Paulo, S\~ao Paulo, Brazil}
\author{  S. K. Adhikari\footnote{Email: adhikari44@yahoo.com, URL: http://www.ift.unesp.br/users/adhikari/}}
\affiliation{Instituto de F\'{\i}sica Te\'orica, UNESP - Universidade Estadual Paulista, 01.140-070 S\~ao Paulo, S\~ao Paulo, Brazil}
\author{P. Muruganandam\footnote{Email: murganand@gmail.com}}
\affiliation{Instituto de F\'{\i}sica Te\'orica, UNESP - Universidade Estadual Paulista, 01.140-070 S\~ao Paulo, S\~ao Paulo, Brazil}
\affiliation{School of Physics, Bharathidasan University, 
Palkalaiperur Campus, 
Tiruchirappalli 620024,   
Tamilnadu, 
India}

\begin{abstract}

{A uniform dilute Bose gas of known density has a universal behavior as 
the atomic scattering length tends to infinity at unitarity while most 
of its properties are determined by a universal parameter $\xi$ relating 
the energies of the noninteracting and unitary gases. The usual 
mean-field equation is not valid in this limit and beyond mean-field 
corrections become important. We use a dynamical model including such 
corrections to investigate a trapped disk-shaped dipolar Bose-Einstein 
condensate (BEC) and a dipolar BEC vortex for large scattering length. 
We study the sensitivity of our results on the parameter $\xi$ and 
discuss the possibility of extracting the value of this parameter from 
experimental observables. }
 
 \end{abstract}


\maketitle

\section{Introduction}

The properties of a uniform dilute interacting $-$ Bose or Fermi $-$ atomic 
quantum gas interacting by an $S$-wave contact inteaction
 at zero temperature is determined by 
two scales   $-$ the atomic scattering length $a$ and density $\bar n$. 
As $a \to   \infty$ at unitarity, the first scale is not of 
concern and the observables of the gas are solely determined by density and 
the gas exhibits a universal behavior. Bulk chemical potential $\bar \mu$ of the unitary  gas is 
proportional to the {\it only} available energy scale $-$ the Fermi energy {(or the chemical potential of the noninteracting gas)}
$E_F=\hbar^2(6\pi^2\bar n)^{2/3}/2m$ $-$ so that $\bar \mu = \xi E_F$,
where $\xi$ is a universal parameter and $m$ is the mass of an atom \cite{rmp,stoof,ska,skab,feruni1,feruni2,feruni3,feruni4,feruni5,feruni6,feruni7,feruni8,bosuni1,bosuni2,bosuni3,bosuni4}. 
{Similarly, the energy per particle $\bar E$
of the unitary gas  is proportional to the energy per particle  $E (\equiv 3E_F/5)$
of the noninteracting gas: $\bar E= \xi E$ \cite{feruni1,feruni2}.} 
The Fermi energy 
is a physically meaningful quantity for the Fermi gas, but the same   can also be used 
as an energy scale for the Bose gas \cite{stoof,bosuni1,bosuni3}.  The Bose and Fermi gases { behave similarly 
 at unitarity, because  the Bose gas exhibits fermionization.} If this fermionization 
of the Bose gas is absolute, then   $\xi$ should be the same for the 
Bose and  Fermi gases.

The   parameter $\xi$ relating the energy of the noninteracting and unitary gases,  
 has been 
``measured" experimentally from a study at or near unitarity 
of the density  \cite{rmp,feruni3,feruni4,feruni5}, or of  ground-state energy,
or of sound velocity  \cite{feruni6} of a trapped Fermi gas. But a 
similar experiment is more difficult  for a   Bose-Einstein condensate (BEC) due to 
{a large probability of three-body loss by 
 molecule formation 
 at or near unitarity, which is the threshold for molecule 
formation \cite{bosuni1}.
In the weak-coupling limit ($an^{1/3}< 1$)  the mean-field Gross-Pitaevskii (GP) equation gives  a good
description of a trapped BEC,  where $n$ is the   density. In the strong-coupling 
regime  ($an^{1/3}> 1$) the GP equation highly overestimates the atomic contact interaction and leads to unphysical results.}
Experimental activities to access the strong-coupling regime of a BEC, to test the 
beyond mean-field corrections \cite{jila,rice}, and to extract the parameter $\xi$ from these studies 
have just began \cite{bosuni1}. 
Although unitarity is also the threshold for molecule  formation in
a two-component (spin up and down) 
Fermi gas, the probability of formation of diatomic molecules is highly suppressed in this case
due to Pauli repulsion among spin-parallel fermions in the three-fermion system and hence is not of concern \cite{rmp}.

Lately, BECs of $^{52}$Cr \cite{cr1,cr2} and $^{164}$Dy \cite{dy1,dy2}   with  large dipolar interaction have been observed 
and studied. {The inter-atomic interaction now has two components: an $S$-wave contact interaction and an anisotropic long-range 
dipolar interaction.}
This allows to study the dipolar BEC  with a variable contact interaction \cite{cr1,dy2} using a Feshbach resonance \cite{fesh}. 
The intrinsically anisotropic dipolar BEC \cite{pla}
has many distinct features \cite{cr1,cr2,cr3,cr4,cr5,adhisol}. The stability of a dipolar BEC depends not only on the scattering length, 
but also on the trap geometry \cite{cr1,cr3,cr5}. A disk-shaped trap leads to  a repulsive dipolar interaction and the dipolar BEC 
is more stable, whereas a cigar-shaped trap yields an attractive dipolar interaction and hence favors 
a collapse instability \cite{cr1,cr5,cr6}.

{We  study the static and dynamic properties of a
disk-shaped  dipolar BEC and dipolar BEC vortex,
with the dipole moments aligned perpendicular to the 
plane of the disk,
for large scattering length  using a  
beyond-mean-field model \cite{skab,skab2}
for the BEC-unitarity crossover.
In this paper we consider the strong-coupling limit of the contact interaction only and {\it not} the same limit of dipolar interaction. }
In the weak-coupling limit, this crossover model reduces to the GP equation and the Lee-Huang-Yang (LHY) correction  \cite{lhy},
whereas in the strong-coupling regime of large scattering length 
it reduces to the   universal result at unitarity.
{ We find that the radial 
densities  are sensitive to the parameter $\xi$  
in the strong-coupling regime and hence a study of density is expected to 
yield information about this parameter. However, the frequency of oscillation of the dipolar BEC 
is found to be insensitive to this parameter. From a study of vortices in a disk-shaped dipolar BEC
we find that both density and radius of vortex core are sensitive to this parameter in the strong-coupling 
regime. 
}


In the disk configuration, the dipolar interaction is highly 
repulsive, as parallel dipoles arranged in a plane with the dipole moment perpendicular to the plane repeal 
each other \cite{cr1,cr2}. The strongly repulsive dipolar interaction should  reduce 
three-body loss by
molecule formation 
in the strong-coupling regime, {as the rate of the reaction $3A\to A_2+A$ should be suppressed in this setting with $A$ 
representing a dipolar atom and $A_2$ a molecule.} Also, as both the contact   and long-range dipolar interactions contribute to
molecule formation, the threshold for molecule formation will be displaced 
from unitarity, specially in strongly dipolar BECs,
   thus creating a new scenario of experiment with a dipolar BEC in  the strong-coupling regime to determine the 
 parameter $\xi$. 

In Sec. II we present the mean-field and beyond-mean-field models to study a dipolar BEC in the weak-
and strong-coupling regimes as well as along the BEC-unitarity crossover as the scattering length $a$ 
is increased. We also 
present a Gaussian variational formulation for its solution at unitarity. 
In Sec. III we present the results of  numerical and variational studies of density, 
root-mean square (rms)
sizes, chemical potential, and frequencies  of radial and axial oscillations of a  disk-shaped dipolar BEC 
and BEC vortex. Finally in Sec. IV we present a brief summary and conclusion.

\section{Analytical  Consideration}

\subsection{Dipolar Gross-Pitaevskii Equation}

We consider  a   disk-shaped 
 dipolar BEC  of $N$ atoms, each of mass $m$,  
 using the   GP 
equation  \cite{cr1}
\begin{eqnarray}  \label{gp3d} 
i \frac{\partial \phi({\bf r},t)}{\partial t}
& =&  \biggr[ -\frac{\nabla^2}{2} + V  +\bar \mu(a,N)+
F(a_{dd},N)\biggr] \phi({\bf r},t),
\nonumber \\
\end{eqnarray} 
with the bulk chemical potential 
\begin{equation}\label{gpbulk}
\bar \mu(a,N)=4\pi a n,
\end{equation}
and density $n=N|\phi|^2$.
Here the dipolar nonlinearity 
$ F(a_{dd},N)=N \int U_{dd}({\bf r -r'})|\phi({\bf r'},t)|^2d
{\bf r'}$, 
\begin{equation}\label{pot}
V=\frac{1}{2}(x^2+y^2+\lambda^2 z^2)
\end{equation}
is the harmonic trap,
 ${\bf r}\equiv \{x,y,z\} \equiv \{\rho,z \}$,   
$ U_{dd}({\bf R}) =  
3a_{dd}(1-3\cos^2\theta)/R^3,   $
 ${\bf R=r-r'},$  
 normalization $\int \phi({\bf r})^2 d {\bf r}$ = 1,
    $\theta$ 
the angle between $\bf R$ and the polarization direction   $z$, 
$\lambda\gg 1$  the trap anisotropy,
 $a_{dd}
=\mu_0	\tilde \mu^2 m /(12\pi \hbar^2)$ 
the strength of 
dipolar interaction,  $\mu_0$ 
the permeability of free space, and
 $\tilde \mu$ the (magnetic) dipole moment. 
{ In Eq. (\ref{gp3d}) the ${\bf r}$ and $t$ dependence of $\bar \mu$ and $F$
are not explicitly shown} and  
length is measured in 
units of  $l_0 \equiv  \sqrt{\hbar/m\omega}$, where $\omega$ is the harmonic trap frequency in $x$ or $y$ directions,
 time $t$ in units of $t_0 = \omega^{-1}$. At unitarity,
the bulk chemical potential  of Eq. (\ref{gp3d}) is independent of $a$ and is \cite{stoof}
\begin{equation}\label{unibulk}
\bar \mu(a,N)=\frac{1}{2}\xi(6\pi^2 n)^{2/3}. 
\end{equation}

To obtain a quantized vortex of unit angular momentum $\hbar$; 
around $z$ axis,  we
introduce a phase (equal to the azimuthal angle) in the 
wave function \cite{vortex2}. This
procedure introduces a centrifugal term $1/[2(x^2+y^2)]$ in the potential of the 
GP equation   so that 
\begin{equation}\label{potv}
V=\frac{1}{2}(x^2+y^2+\lambda^2 z^2)+\frac{1}{2(x^2+y^2)}.
\end{equation}
 We
adopt this procedure to study an axially-symmetric vortex in a disk-shaped dipolar
BEC.

\subsection{BEC-Unitarity Crossover}

Lee,   Huang, and Yang (LHY) \cite{lhy} obtained the leading terms of the beyond-mean-field 
expression for energy of a uniform Bose gas from which the following expression for 
 the bulk chemical 
potential can be obtained
\cite{fp}:
\begin{equation}\label{lhy}
\bar \mu(a,N)= {4\pi   an}  \left[{1+\alpha a^{3/2}\sqrt{n} }     \right].
\end{equation}
The lowest order term in this 
expansion is the GP result (\ref{gpbulk})
first derived by Lenz \cite{lhy3}. However, expression (\ref{lhy}), although gives the leading correction for 
larger $an^{1/3}$,  
diverges in the strong-coupling regime,  and
hence has only limited validity along the BEC-unitarity crossover.

In addition to studying the system in the  weak-coupling limit (\ref{gpbulk}) and
unitarity (\ref{unibulk}),
we   also consider the system along the full 
BEC-unitarity crossover from weak to strong coupling,
as the parameter  $an^{1/3}$ is increased.
For this purpose we consider the following 
minimal crossover model  for 
the bulk chemical potential   consistent with weak and 
strong couplings \cite{skab,ska}
\begin{eqnarray}\label{bulk}
&&\bar \mu(a,N)= 4\pi  n^{2/3}f(\chi),\quad \chi=an^{1/3}, \\
&&f(\chi)= \left[\frac{\chi+(1+\nu)\alpha \chi^{5/2}}{1+\nu\alpha \chi^ {3/2}
+(1+\nu)\gamma  \chi^{5/2}}     \right],\label{bulk2}
\end{eqnarray}
where 
$\alpha=32/(3\sqrt \pi), \alpha/\gamma= \xi (6\pi^2)^{2/3}/8\pi $, and $\nu$ is the only free parameter in this expression. 
The parameters $\alpha$ and $\gamma$ are determined by the constraints that   expression (\ref{bulk}) be consistent 
with the LHY correction (\ref{lhy}) as well as the unitarity limit (\ref{unibulk}),
   both independent of the parameter $\xi$.     
Expression 
  (\ref{bulk}) is weakly sensitive to $\nu$ and a smooth interpolation between the weak and strong-coupling 
regimes is obtained for any small $\nu$. In this study we   use $\nu =1$. This  value of $\nu$ was used  \cite{ska}
successfully in a study of $^6$Li$_2$ BEC in the BEC-unitarity crossover.   
Expression 
  (\ref{bulk}) can also reproduce fairly well \cite{skab2} the energies of diffusion  Monte Carlo (DMC) calculation  \cite{DMC} of a trapped 
bosonic system of small number of atoms. It was found that the energies obtained from the crossover model (\ref{bulk}) 
for that bosonic system
are in better 
agreement with the DMC calculation than those obtained from the GP equation.

{ The crossover model (\ref{bulk}) is Galilei invariant 
and yields the
hydrodynamic equations of the dipolar BEC at zero
temperature, and enables one to study collective dynamical
properties of the system in the full crossover from
weak-coupling to unitarity \cite{skab,skab2}. 
Equations (\ref{gp3d}) and (\ref{bulk}) should be
 considered as a generalization of the  GP equation 
with beyond mean-field corrections to properly include 
the effect of interaction for large positive scattering length. The saturation 
of the interaction at unitarity is properly taken care of in the crossover model (\ref{bulk}).  
As an application we shall study 
here the properties of a disk-shaped dipolar BEC and dipolar BEC vortex in the strong-coupling 
regime to show the sensitivity of the result to the universal parameter $\xi$.
    }

\subsection{Variational Approximation at Unitarity}

\label{sec:vari}
 
At unitarity, the dipolar mean-field  equations (\ref{gp3d}), (\ref{pot}), and (\ref{unibulk}) can be 
conveniently solved by a 
time-dependent Lagrangian variational approach. This can be used to study the
size and frequencies of oscillation of the dipolar BEC at unitarity.
  This is done by reducing Eq.
(\ref{gp3d}) to a system of second order nonlinear ordinary
differential equations involving the variational parameters.
The Lagrangian density of Eq.  (\ref{gp3d})  is given by \cite{adhisol}
\begin{align}& \,
{\mathcal L}=\frac{i}{2}\left( \phi \phi^{\star}_t
- \phi^{\star}\phi_t \right) +\frac{1}{2}\vert\nabla\phi\vert^2
+ \frac{1}{2}(\rho^2+\lambda^2z^2)
\vert\phi\vert^2 \notag \\ & \,
+ {\frac{3\xi}{10}(6\pi^2N)^{2/3}\vert\phi\vert^{10/3}} + \frac{N}{2}\vert\phi\vert^2\int U_{dd}({\mathbf r}-
{\mathbf r'})\vert\phi({\mathbf r'})\vert^2 d{\mathbf r}' .\label{eqn:vari}
\end{align}
{ Recalling that $n=N |\phi|^2$, it can be straightforwardly verified that Eqs. (\ref{gp3d}) and (\ref{unibulk}) are the Euler-Lagrange 
equations for the Lagrangian density (\ref{eqn:vari}) \cite{adhix}, which should be used in the variational formulation 
\cite{perez}. 
}
To develop the variational approximation, we consider the following Gaussian 
ansatz for the wave function \cite{adhisol}
\begin{align}
\phi({\mathbf r}, t) = \sqrt{ \frac{\pi^{-\frac{3}{2}}}{w_\rho^2 w_z} }
 \, \exp\left[-\frac{\rho^2}{2w_\rho^2}-\frac{z^2}{2w_z^2}
+i\alpha \rho^2+i\delta z^2 \right], \label{eq:trial}
\end{align}
where the time-dependent variational parameters $w_\rho$ and
$w_z$ are the  radial and axial
widths and  $\alpha$ and $\delta$ are the chirps.
The Lagrangian density can be calculated   by substituting
the   wave function (\ref{eq:trial}) in
Eq. (\ref{eqn:vari}). Then the
effective Lagrangian $L\equiv  \int {\mathcal L}\,d{\mathbf r}$ becomes
\begin{align}
&\,
L  
 =  \frac{1}{2}\left({ 2}w_\rho^2\dot{\alpha} +
w_z^2\dot{\delta}\right) +
\frac{1}{2}\bigg(\frac{1}{w_\rho^2} + \frac{1}{2w_z^2}
+ { 4}w_\rho^2 \alpha^2 + 2w_z^2\delta^2\bigg) \notag \\ & \,
+ \frac{1 }{4}
\left(2 w_\rho^2 + \lambda^2 w_z^2
\right) 
- \frac{N a_{dd}}{\sqrt{2 \pi}}
\frac{f(\kappa)}{w_\rho^2w_z}  
+ \frac{9{\cal C}}{w_z^{2/3}w_\rho^{4/3}}
, \label{lag:eff}
\end{align}
%
where ${\cal C}= \sqrt{3}\xi (6\pi^2N)^{2/3}/(50\pi \sqrt{5}), \kappa=w_\rho/w_z   $, and   
\begin{align}
f(\kappa)=\frac{1+2\kappa^2}{1-\kappa^2} -\frac{3\kappa^2\mbox{tanh}^{-1}
\sqrt{1-\kappa^2}}{(1-\kappa^2)^{\frac{3}{2}}}. \label{eqn:fkappa}
\end{align}

The corresponding Euler-Lagrange equations
governing  the evolution of
the widths $w_\rho$ and $w_z$ yield
\begin{subequations}
\begin{align}
  &
\ddot{w}_{\rho}+  {w_\rho} =
\frac{1}{w_\rho^3} -\frac{
a_{dd}}{\sqrt{2\pi}} \frac{Ng(\kappa)}{w_\rho^3w_{z}}
+ \frac{12 {\cal C}}{w_z^{2/3}w_\rho^{7/3}},
\label{13a} \\ &
 \ddot{w}_{z} +  \lambda^{2} w_{z} =
\frac{1}{w_z^3}- \frac{ a_{dd}}{\sqrt{2\pi}}
\frac{2N}{w_\rho^2w_z^2} 
h(\kappa)+ \frac{12{\cal C}}{w_z^{5/3}w_\rho^{4/3}},
\label{13b}  
\end{align}\end{subequations}
where
\begin{subequations}
\begin{align}
& g(\kappa) = \frac{2-7\kappa^2-4\kappa^4}{(1-\kappa^2)^2} + 
\frac{9\kappa^4\mbox{tanh}^{-1}\sqrt{1-\kappa^2}}{(1-\kappa^2)^\frac{5}{2}}, \\
& h(\kappa) = \frac{1+10\kappa^2-2\kappa^4}{(1-\kappa^2)^2} - 
\frac{9\kappa^2\mbox{tanh}^{-1}\sqrt{1-\kappa^2}}{(1-\kappa^2)^\frac{5}{2}}
.
\end{align}
\end{subequations}

Equations (\ref{13a}) and (\ref{13b})
provide the dynamics  of the evolution of
 radial and axial widths, respectively.
One can obtain the expression for the frequencies and lowest-lying modes
from these equations~\cite{vortex}. The widths for a stationary state can
be obtained by setting
$\ddot w_\rho  =0$ and $\ddot w_z =
0$ in Eqs. (\ref{13a}) and (\ref{13b}).
 The chemical potential $\mu$ for the stationary state is given by 
\begin{align}
&\,
\mu 
 = 
\frac{1}{2}\bigg(\frac{1}{w_\rho^2} + \frac{1}{2w_z^2} \bigg)
+ \frac{1 }{4}
\left(2 w_\rho^2 + \lambda^2 w_z^2
\right) \notag \\ & \,
- 2\frac{N a_{dd}}{\sqrt{2 \pi}}
\frac{f(\kappa)}{w_\rho^2w_z}  
+ \frac{5}{3}\frac{9{\cal C}}{w_z^{2/3}w_\rho^{4/3}}
. \label{lag:eff}
\end{align}

\section{Numerical Calculation}

We perform    numerical simulation of the 3D GP equation (\ref{gp3d})
using   the split-step  Crank-Nicolson 
method  \cite{Muruganandam2009}.  {
The
evaluation of the dipolar integral term in this equation in coordinate space is not straightforward due to the
divergence at short distances.
However, this has been tackled by evaluating the dipolar term in the momentum (k) space. The integral
can be simplified in Fourier space by means of convolution as \cite{cr3}
\begin{equation}\label{con}
\int d{\bf r}'U_{dd}({\bf r-r'})|\phi({\bf r'})|^2= {\cal F}^{-1}\{ {\cal F}[U_{dd}]({\bf k}) {\cal F}[|\phi|^2]({\bf k}) \}({\bf r}),
\end{equation}
where ${\cal F}[$ $]$ and ${\cal F}^{-1}\{ \}$ are the Fourier transform (FT) and inverse FT, respectively.   The FT
of the dipole potential is  known analytically \cite{cr3}.
The FT of  density $|\phi|^2$ is evaluated numerically by means of a standard fast FT (FFT) algorithm. 
The dipolar integral in Eq. (\ref{gp3d}) involving the FT of density multiplied by FT
of dipolar interaction is evaluated by  the convolution theorem (\ref{con}). The inverse FT is taken by means
of a standard FFT algorithm. The FFT algorithm is carried out in Cartesian coordinates and hence the GP
equation is solved in  three dimensions irrespective of the symmetry of the trapping potential.
In the Crank-Nicolson algorithm we used space step 0.1, time step 0.002 and employed upto 512 space discretization points
in each Cartesian direction. We made an error analysis of the results for chemical potential and rms sizes and found that 
the maximum numerical    error of the results reported here
is less than 0.5 $\%$.   }

\subsection{Experimental Considerations}

Of the experimental dipolar BECs $-$ $^{52}$Cr and $^{164}$Dy $-$ realized so far, the magnetic moment 
of  $^{52}$Cr is $\tilde \mu = 6\mu_B$ \cite{cr1}, where $\mu_B$ is the Bohr magneton, and that of $^{164}$Dy is 
$\tilde \mu = 10 \mu_B$ \cite{dy2}. Consequently, $a_{dd}\equiv \mu_0\tilde \mu^2 m /(12\pi \hbar^2)= 15a_0$ for $^{52}$Cr
and $a_{dd}= 130a_0$ for   $^{164}$Dy, with $a_0$ the Bohr radius.
 Hence the dipolar interaction in $^{164}$Dy is about 9 times stronger than in  $^{52}$Cr and 
we  employ a $^{164}$Dy BEC in this study. 
For $^{164}$Dy,  an estimate for the scattering length is 
$a\approx 100a_0$ \cite{dy2}.  In the actual 
experiment on $^{164}$Dy a dipolar BEC of 15000 atoms in
a fully anisotropic 
trap with  frequencies  $\{f_x,f_y,f_z\}=\{380,500,1570\}$ Hz  was obtained \cite{dy2}.
In this study, to simulate this experiment \cite{dy2},  we use 
the frequencies $\{f_x,f_y,f_z\}=\{436,436, 1570\} $ Hz, so that 
$\lambda =3.601$,  where we take a geometrical mean of the 
frequencies in $x$ and $y$ directions  to generate an axially-symmetric BEC.  
The length scale employed here, for $\omega=2\pi \times 436$ Hz,  is  $l_0=\sqrt{\hbar/(m\omega )} = 0.376$ $\mu$m.

\begin{table}
\caption{Theoretical result and experimental evaluation of the parameter 
$\beta\equiv (\xi-1)$ for the Bose and Fermi gases. }
 
\centering
\begin{tabular}{lrc}
\hline
\hline
Fermi, Theory&  Astrakharchik {\it et al.}\cite{feruni1}&  $-0.58$  \\
& Carlson {\it et al.} \cite{feruni2}&  $-0.58$  \\
       & Perali {\it et al.} \cite{perali}&   $-0.545$  \\
\hline 
Fermi, Expt ($^6$Li) & Partridge {\it et al.} \cite{feruni3}& $-0.54(5)$\\
                     &Kinast {\it et al.} \cite{feruni4}& $-0.49(4)$\\
                      &Bartenstein {\it et al.} \cite{feruni5}  & $  -0.73^{+0.12}_{-0.09}$\\
                &Navon {\it et al.} \cite{feruni8} &$ -0.59(1)$ \\
             &Luo and Thomas \cite{feruni6} &$\approx  -0.6$ \\
\hline
    Fermi, Expt ($^{40}$K)        &Stewart {\it et al.} \cite{feruni7}&$-0.54 ^{+0.05}_{-0.12}$\\
\hline
Bose, Theory&  Diederix {\it et al.} \cite{stoof} & $-0.54$ \\
            &Lee {\it et al.} \cite{bosuni4} & $-0.34$ \\
            &Cowell {\it et al.} \cite{bosuni2}& $<1.93$ \\
                 &        Song and Zhou \cite{bosuni3}          & $<-0.2$\\ 
Analysis, Expt ($^6$Li$_2$) \cite{feruni5}& Adhikari \cite{ska}& $<0.6$ \\
\hline
Bose, Expt ($^7$Li) & Navon {\it et al.} \cite{bosuni1}& $>-0.56(8)$ \\
\hline
\end{tabular}
\label{I}
\end{table}

Next we summarize the different 
theoretical and 
experimental estimates of $\xi$ obtained so far for bosons and fermions.
Often the parameter $\xi $ is written as $\xi \equiv (1+\beta)$ and different 
estimates of
 $\beta$ is given in Table \ref{I}, where the variational calculations of Refs. \cite{bosuni2,bosuni4} for bosons are 
upper bounds and the experimental result of Ref. \cite{bosuni1} for $^7$Li is a lower bound only.
Yet another estimate of $\xi$ can be obtained from a consideration of Fermi superfluid in the BEC side 
of the Bardeen-Cooper-Schrieffer-BEC (BCS-BEC) crossover. Here we reconsider an analysis \cite{ska} of 
the   experiment \cite{feruni5} on  $^6$Li    in the BEC 
side of BEC-unitarity crossover.
The molecular BEC of $^6$Li$_2$
was then studied using Eqs.
(\ref{gp3d}), (\ref{bulk}), and (\ref{bulk2}) but with $\alpha/\gamma= \xi_{	\text{mol}} (6 \pi^2)^{2/3}/(2\pi)$,
where $\xi_{	\text{mol}}$ is the universal parameter of Ref. \cite{ska},
in place of $   \alpha/\gamma= \xi (6 \pi^2)^{2/3}/(8\pi)$ considered here. This implies that for a comparison of the 
two studies we should take $\xi=4\xi_{	\text{mol}}$. The analysis of Ref. \cite{ska} yielded $\xi_{	\text{mol}}
\approx 0.4$, so that   $\xi \approx 1.6$ corresponding to $\beta \approx 0.6$.  
In that analysis \cite{ska} it was assumed that 
the bosonic molecular unitarity of $^6$Li$_2$ was achieved 
for the same strength of atomic interaction  as the fermionic unitarity of  $^6$Li. 
Actually, the bosonic 
molecular unitarity should be achieved at a different value of interaction and into the BEC side of 
the BCS-BEC 
crossover,  where the system is less repulsive. This would lead to a smaller value of the parameter $\xi$ ($\xi<1.6$) and $\beta$.
Hence the analysis of Ref. \cite{ska} gives  an upper bound. 
From the results reported in Table \ref{I}, 
the most accurate theoretical \cite{feruni1,feruni2} and experimental \cite{feruni6,feruni8}
estimates for a Fermi gas converge to a value of $\xi$ very close to 0.4. 

\subsection{Disk-shaped dipolar Bose-Einstein condensate}

We study a disk-shaped dipolar BEC of 15000 $^{164}$Dy atoms with 
$a_{dd}=130a_0$
as in the experiment 
of Lu {\it et al.} \cite{dy2}.
{ The parameter $\xi$ can be extracted from the observables of the dipolar BEC 
in the strong-coupling regime where the observables would be sensitive to this parameter. 
For this 
purpose, in this paper, in addition to the numerical 
study at unitarity, we also present a complete numerical 
study of the dipolar BEC in the strong-coupling regime for 
  scattering length $a> 100a_0$ using the crossover model (\ref{bulk}). 
}

\begin{figure}[!t]
\begin{center}
\includegraphics[width=\linewidth,clip]{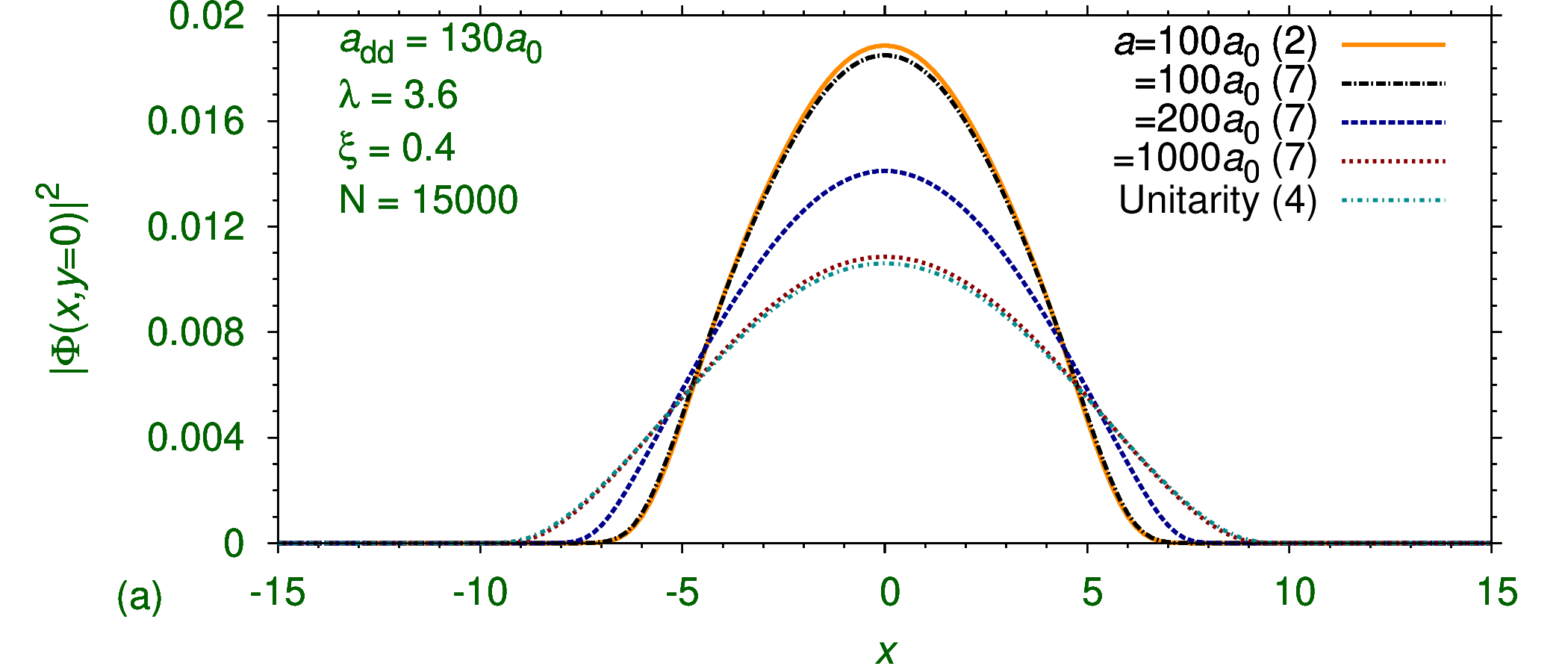}
\includegraphics[width=\linewidth,clip]{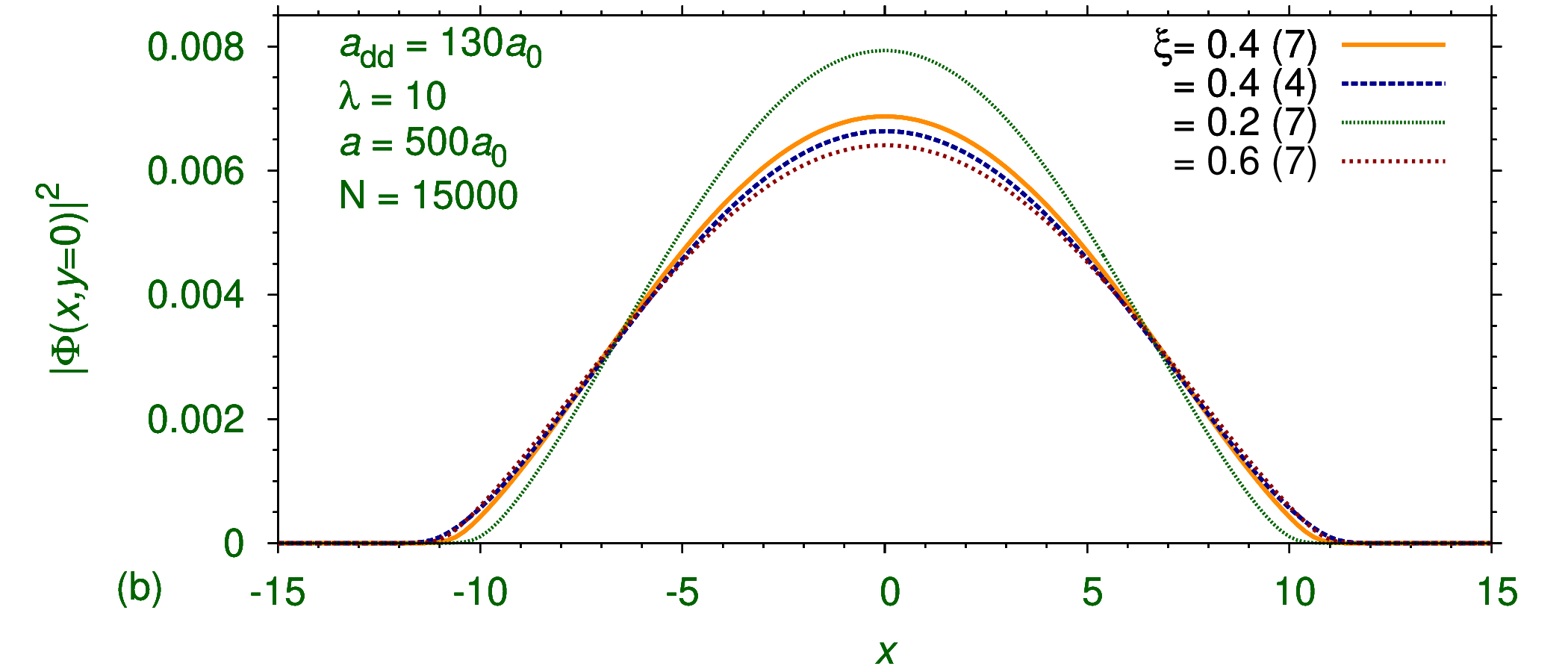}
\includegraphics[width=\linewidth,clip]{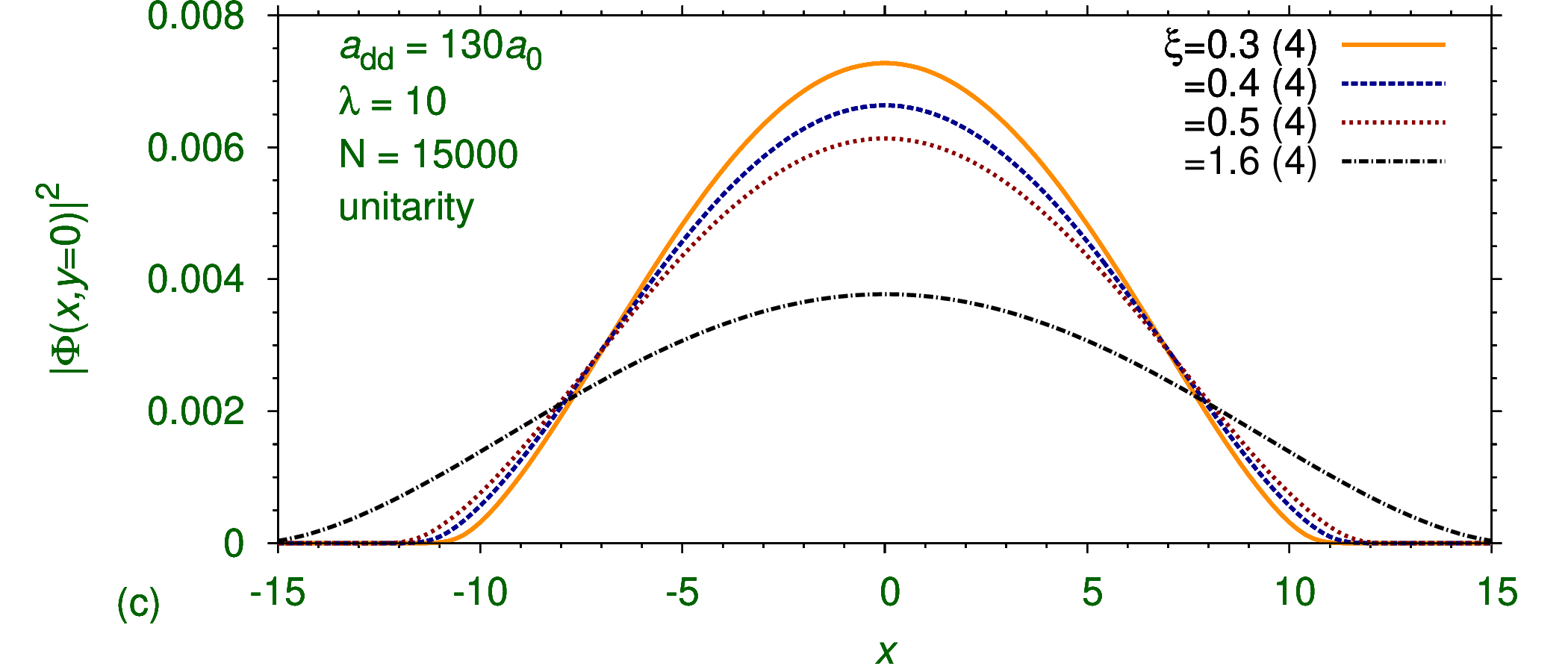}
\includegraphics[width=\linewidth,clip]{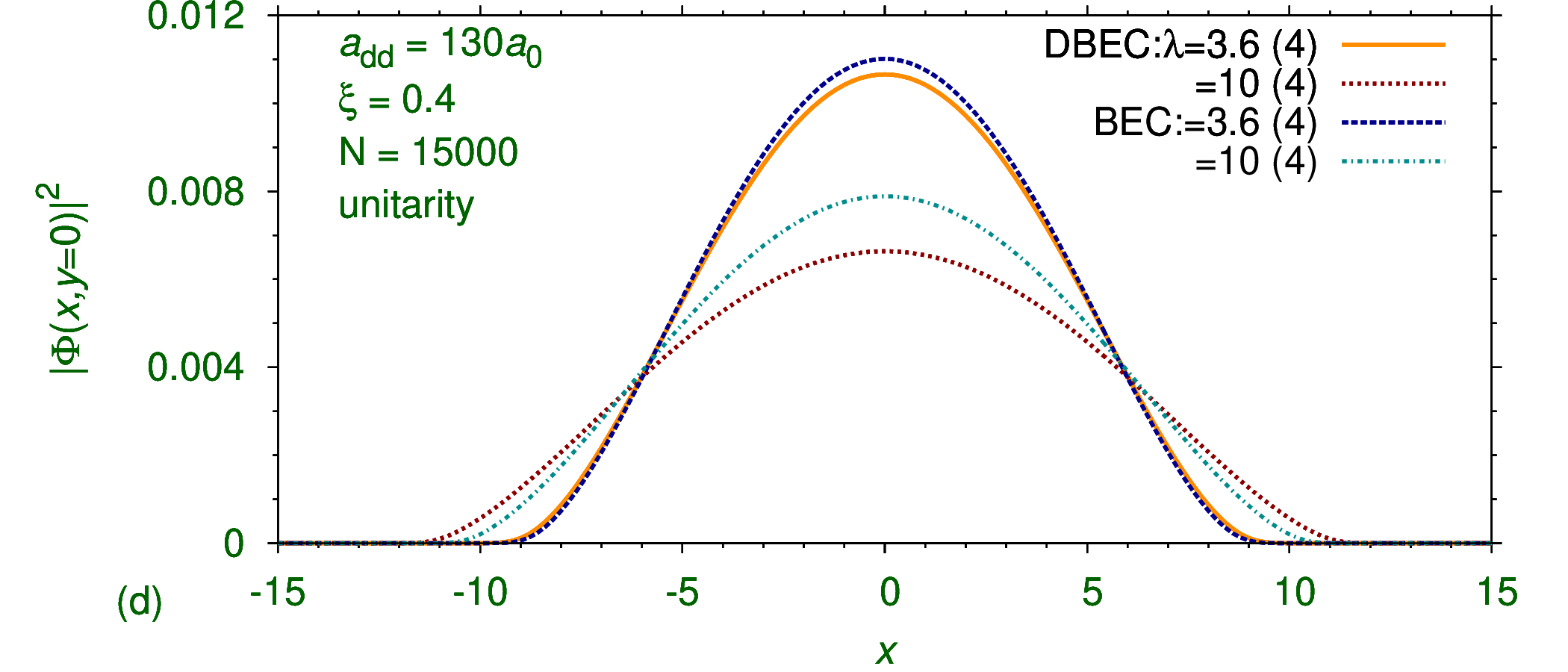}
\end{center}

\caption{(Color online) (a) Radial density along $x$ axis $|\Phi(x,y=0)|^2=
\int dz |\phi(x,y=0,z)|^2$ 
of a $^{164}$Dy BEC with $N=15000$, $\lambda
=3.6$, $a_{dd}=130a_0, \nu=1$ and $\xi =0.4$ for  $a=100a_0,
200a_0, 1000a_0$ and at unitarity  using 
the GP limit (\ref{gpbulk}), crossover model (\ref{bulk}), and at unitarity 
(\ref{unibulk}). 
(b) The same  for  $\lambda
=10$,   $a=500a_0$ and $\xi = 0.2,0.4,0.6$   using the crossover model 
(\ref{bulk}). The density at unitarity  
(\ref{unibulk}) for $\xi=0.4$ is also shown. 
(c) The same for  $\lambda
=10$ and different 
$\xi$ at unitarity  (\ref{unibulk}).
 (d)  The same for  $\xi =0.4, \lambda = 3.6,10$ at unitarity  (\ref{unibulk}) for $a_{dd}=130a_0$ (DBEC), and $a_{dd}=0$ (BEC).}
  
\label{fig1}
\end{figure}

 In Fig. \ref{fig1} (a), we plot the radial 
density of the BEC along  $x$ axis $|\Phi(x,y=0)|^2$ 
obtained by integrating out the $z$ dependence 
of density: $|\Phi(x,y)|^2=\int dz |\phi(x,y,z)|^2$. 
In this figure we show the 
result for $a=100a_0$  using the   GP 
model (\ref{gpbulk}) and  at unitarity (\ref{unibulk}) in addition to the results  
for $a=100a_0, 300a_0,$ and $500a_0$ using the BEC-unitarity 
crossover model (\ref{bulk}) with $\xi =0.4$ and $\nu=1$.  For small $a$, the 
density from
 the crossover model (\ref{bulk})
is in agreement with the GP model (\ref{gpbulk})
 and hence practically independent of the parameter $\xi$, whereas 
 for large $a$ it  
approximates the unitarity limit (\ref{unibulk}) with the increase of $a$. 
 In Fig. \ref{fig1} (b) we plot the 
radial density for $a=500a_0$ for different $\xi$ using the crossover model (\ref{bulk}). 
The result  at unitarity (\ref{unibulk}) for $\xi=0.4$ is also shown. 
The density is sensitive to the parameter $\xi$ for $a=500a_0$ as can be seen from Fig. 
\ref{fig1} (b) comparing the results for $\xi=0.2,0.4$ and 0.6. The sensitivity of the 
density on $\xi$ at unitarity is illustrated 
in Fig. \ref{fig1} (c), where we show the radial density for
$\xi =0.3,0.4, 0.5$ and 1.6. 
Finally, in Fig. \ref{fig1} (d) we show the density at unitarity for nondipolar and dipolar BECs
for two values of the trap asymmetry $\lambda = 3.6 $ and 10. The difference between the two densities 
is more pronounced for $\lambda =10$, where the dipolar repulsion is stronger.

\begin{figure}
\begin{center}
\includegraphics[width=\linewidth,clip]{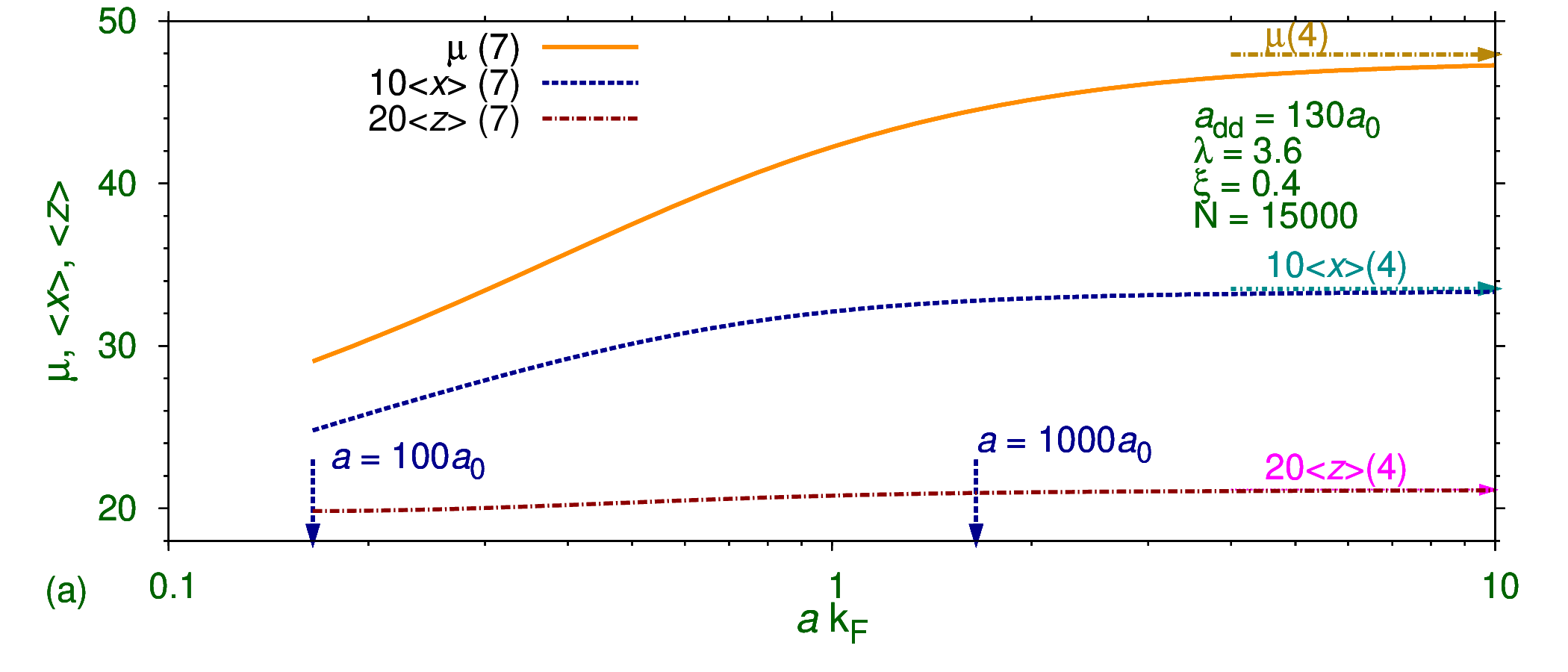}
\includegraphics[width=\linewidth,clip]{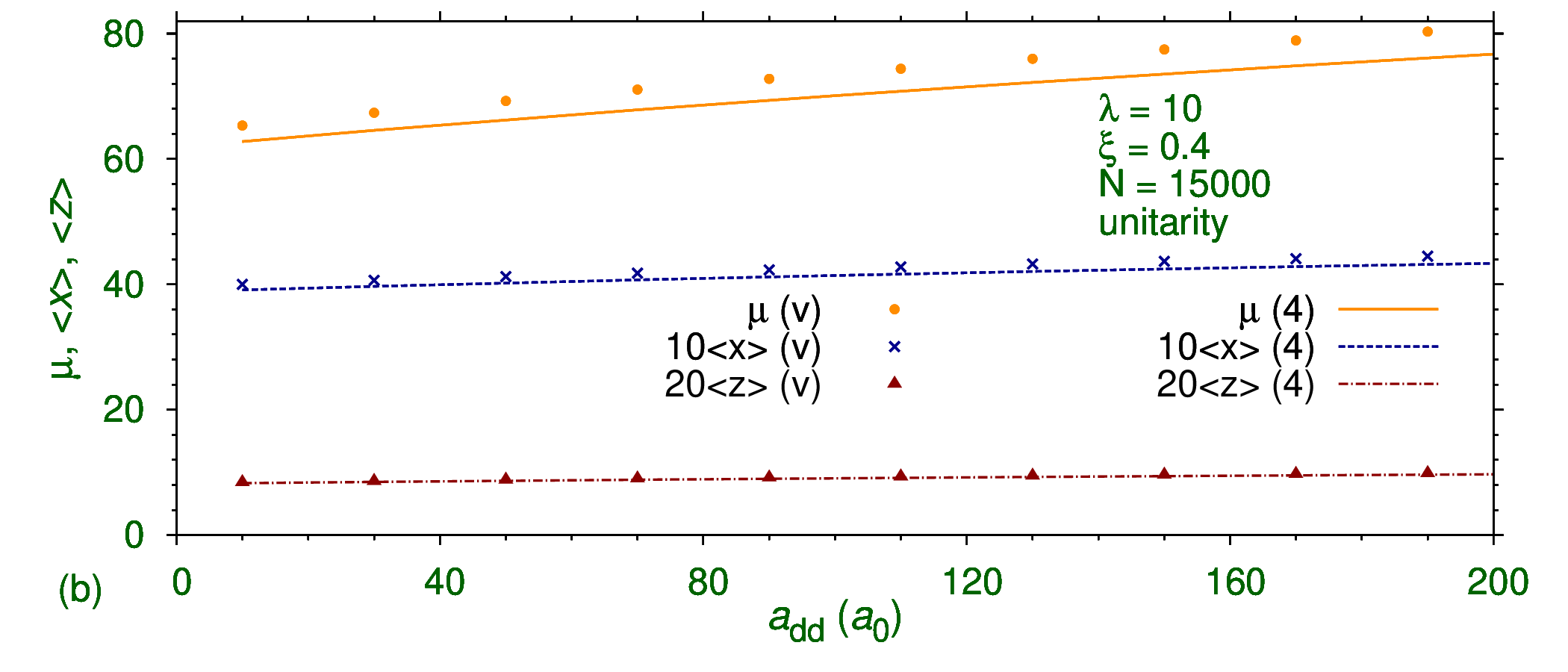}
\includegraphics[width=\linewidth,clip]{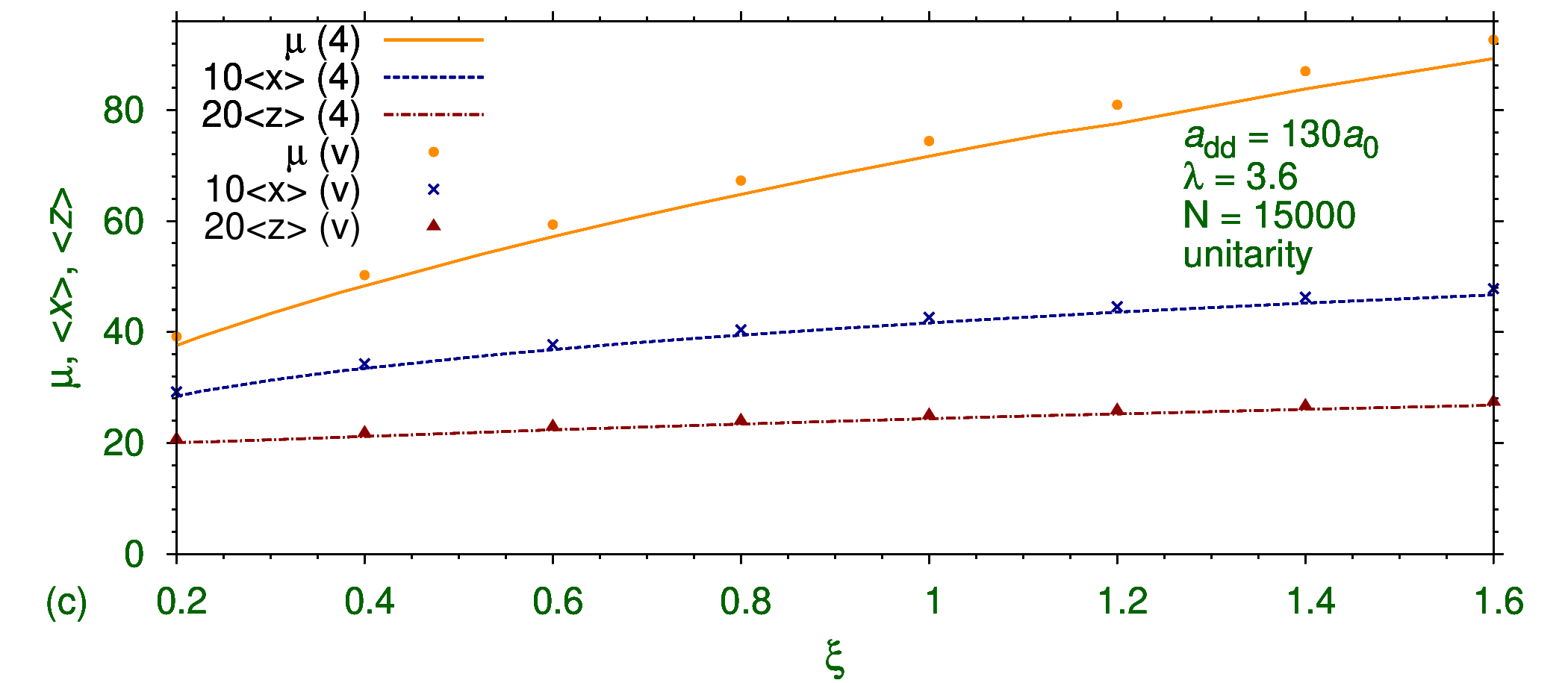}
\includegraphics[width=\linewidth,clip]{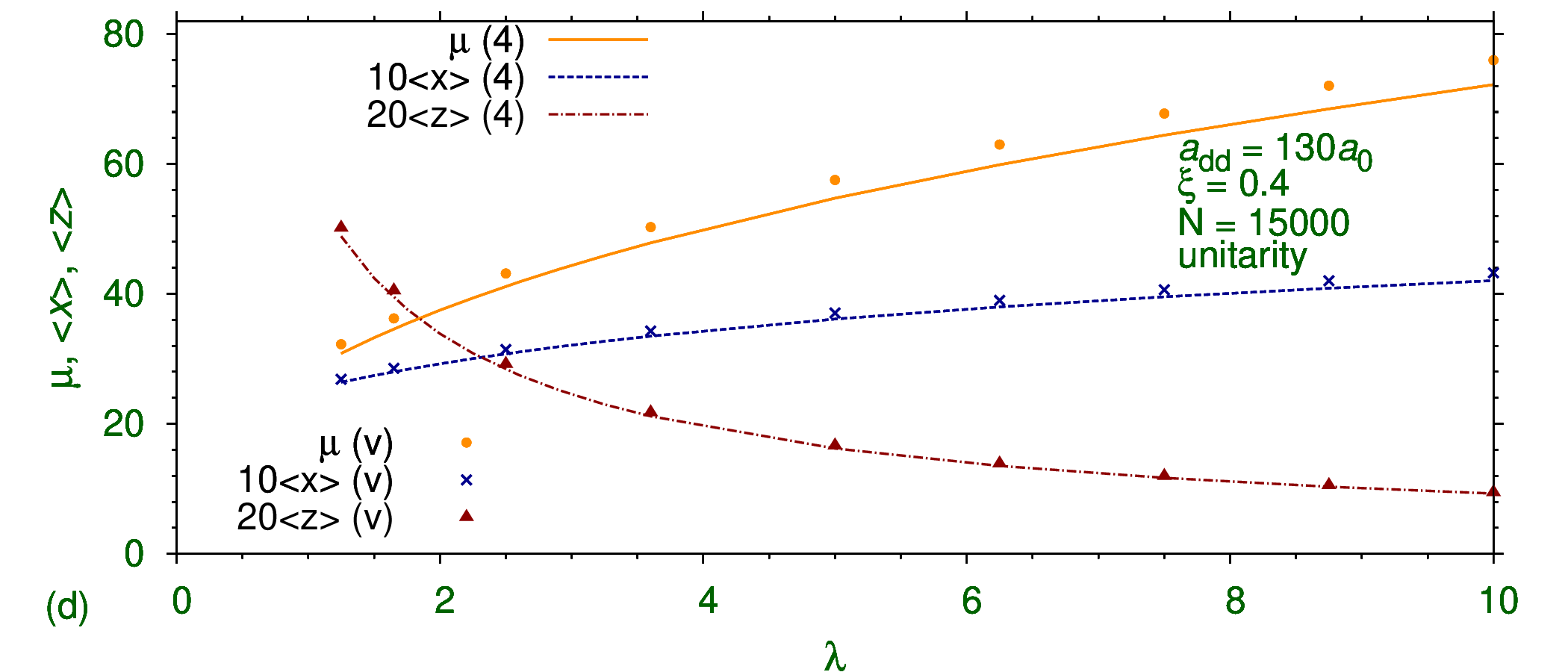}
\end{center}

\caption{(Color online) (a) The numerical   rms sizes $\langle x \rangle$, $\langle z \rangle$, 
and chemical potential $\mu$ of the dipolar BEC with $N=15000, a_{dd}
=130a_0, \xi=0.4, \nu=1, \lambda =3.6$ versus 
$ak_F$ using the   crossover model 
(\ref{bulk}),  as well as  at unitarity (\ref{unibulk}) (arrow).
(b) The numerical (\ref{unibulk})
and variational (v) results for the same  at unitarity 
versus $a_{dd}$. 
(c) The numerical (\ref{unibulk})
and variational (v) results for the same  at unitarity 
versus $\xi$. 
(d) The numerical (\ref{unibulk})
and variational (v) results for the same  at unitarity 
versus $\lambda$.  
}
\label{fig2}
\end{figure}

In Fig. \ref{fig2} (a) we plot the rms sizes of the BEC
$\langle x \rangle$, $\langle z \rangle$, and the chemical potential $\mu$
for $\xi=0.4,$ and $ a_{dd}=130a_0$ as calculated using crossover model (\ref{bulk}),
versus the dimensionless 
parameter $k_Fa$, where $k_F$ is the  Fermi wave vector in a harmonic trap 
defined by 
$k_F=  (48N)^{1/6}/\bar l$ \cite{rmp}, where $\bar l= \sqrt
{\hbar/m \bar \omega}, \bar \omega = 2\pi(f_x f_y f_z)^{1/3}$, $\{f_x, f_y, f_z\} = \{436, 436, 1570\}$ Hz.  
 One can find from Fig. \ref{fig2} (a)
how these quantities $-$ $\langle x \rangle$, $\langle z \rangle$, $\mu$
$-$ approach their values at unitarity as $a$ increases. With the increase of $a$ these
quantities saturate rapidly to their respective values at unitarity. 
In Fig. \ref{fig2} (b),  we plot the numerical (n)
and variational (v) results for 
$\langle x \rangle$, $\langle z \rangle$, and
$\mu$ versus $a_{dd}$ at unitarity, which shows that the results are weakly sensitive to a variation of $a_{dd}$.
 In Fig.  \ref{fig2} (c) we plot $\langle x \rangle$, $\langle z \rangle$, and $\mu$ versus $\xi$ at unitarity, 
which shows that the results are   sensitive to a variation of $\xi$. Finally, in 
Fig.  \ref{fig2} (d) we plot
$\langle x \rangle$, $\langle z \rangle$, and $\mu$ versus $\lambda$ at unitarity showing the strong sensitivity of 
the results to a variation of $\lambda$. From Figs. \ref{fig2} (b), (c), (d) we see that the variational 
results are in good agreement with the numerical ones.

In a recent experiment, Navon {\it et al.} \cite{bosuni1} 
were able to make measurements for densities of a very dilute BEC of  $^7$Li
for   $a=2150a_0$
and extract the parameter $\xi$ from a theoretical analysis using the LHY correction \cite{lhy,fp}.
The very dilute BEC prepared in a weak trap, 
 allowed to make an experiment for  large $a=2150a_0$.
But because of the low density, the BEC remained away from the strong-coupling regime even for $a=2150a_0$
and was studied by the LHY correction,  rather than a full crossover model as in the present study. 
In this regime the densities are weakly sensitive to the parameter $\xi$
and only an upper limit $\xi<0.6$  could be obtained from that study \cite{bosuni1}. 

\begin{figure}
\begin{center}
\includegraphics[width=\linewidth,clip]{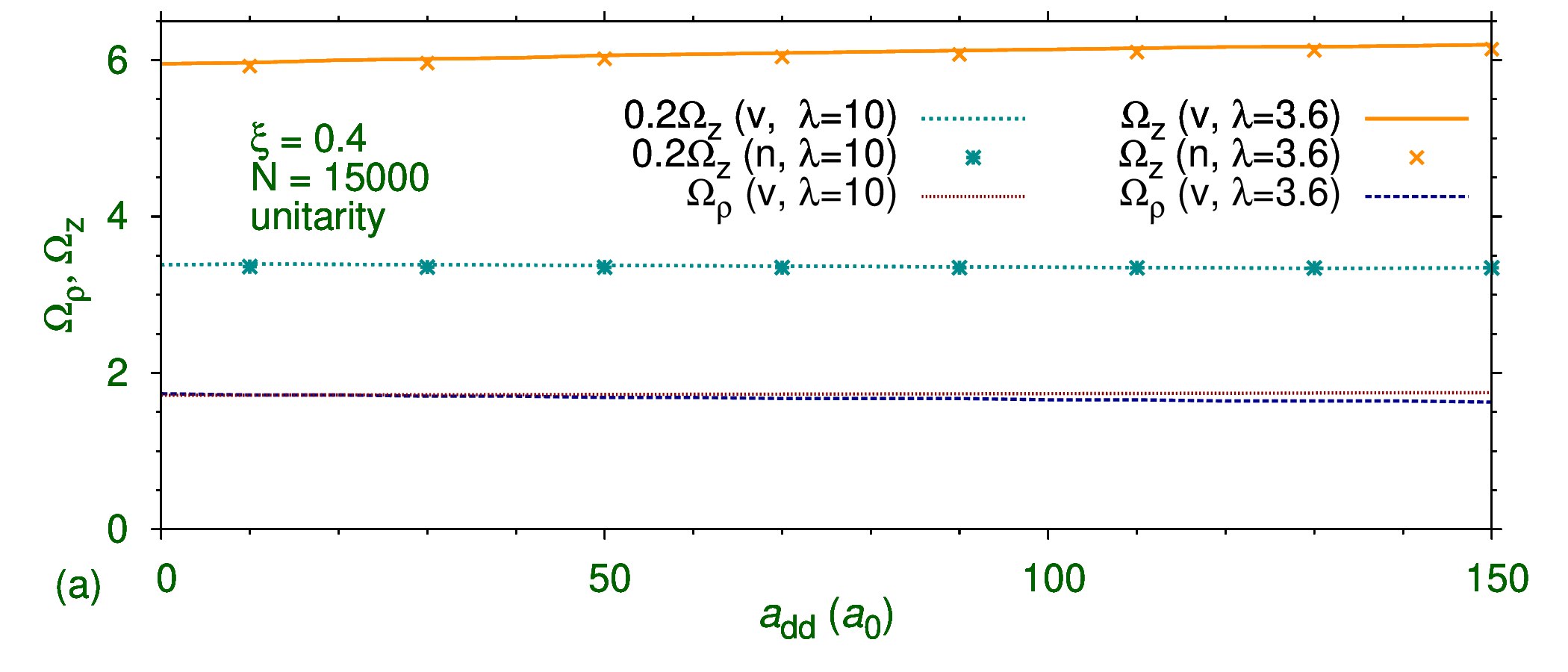}
\includegraphics[width=\linewidth,clip]{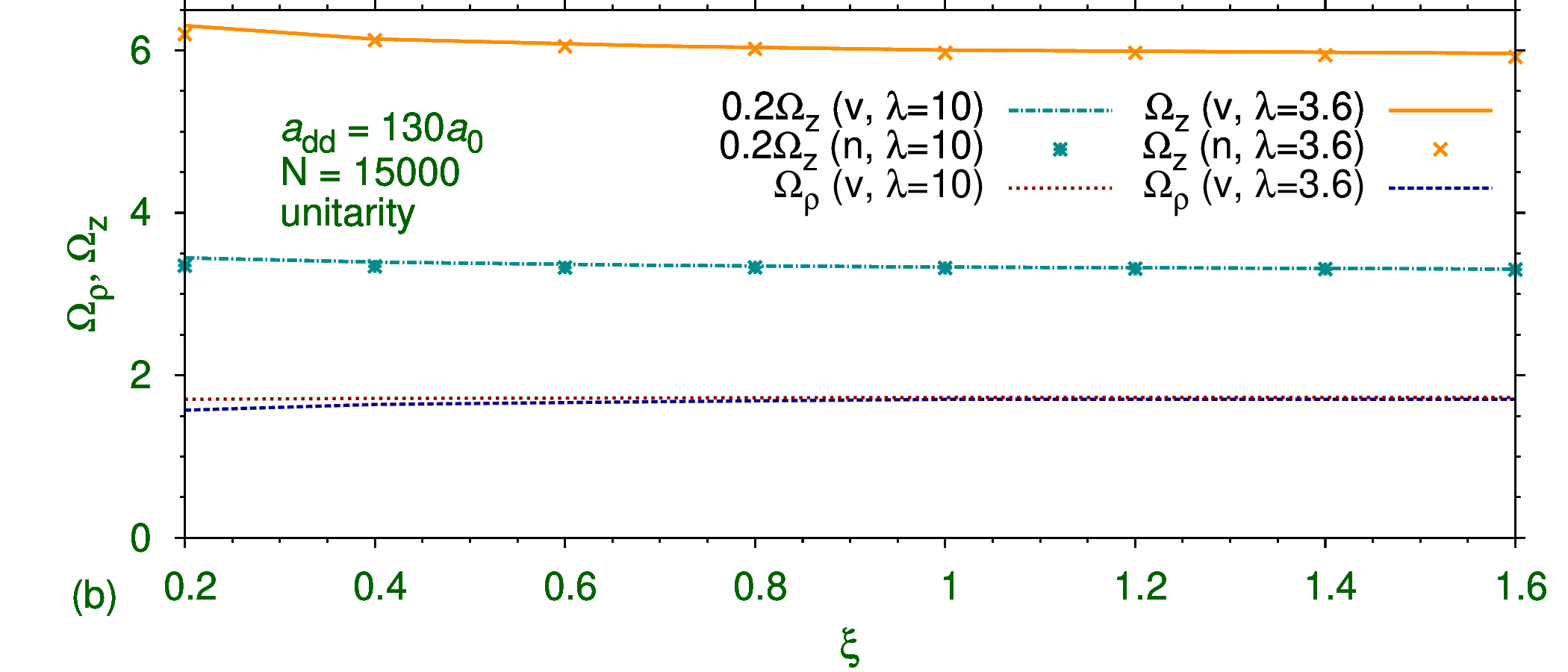}
\end{center}

\caption{(Color online) (a) Variational frequency (v) of radial and axial oscillation $\Omega_\rho$ and $\Omega_z$
 of the dipolar BEC with $N=15000,  \xi=0.4, \nu=1, \lambda =3.6,10 $ versus 
$a_{dd}$ at unitarity from Eqs. (\ref{13a}) and (\ref{13b}).
The numerical frequencies (n) of axial oscillation  obtained from a solution of Eqs. (\ref{gp3d}) and (\ref{bulk}) are also shown. 
(b) The same for $a_{dd}=130a_0$ versus $\xi$. 
}

\label{fig3}
\end{figure}

Apart from density profile and rms sizes of the dipolar BEC, other observables which can be  
studied are the   frequencies of radial and axial oscillations $\Omega_\rho$ and $\Omega_z$, respectively,
of the fundamental modes.   
We calculated these frequencies by numerically solving the variational equations 
 (\ref{13a}) and (\ref{13b}) in different cases. {  The initial widths $w_\rho$ and $w_z$ were 
taken as their equilibrium static values and their time evolution is obtained. The frequencies of radial and axial oscillations
were extracted from the time evolution of the respective widths.  } 
In Fig. \ref{fig3} (a) we show these frequencies for $\xi=0.4$ versus $a_{dd}$ and in 
    Fig. \ref{fig3} (b) we plot these frequencies 
for $a_{dd}=130a_0$ versus $\xi$ for $\lambda =3.6, 10$. We also calculated 
the axial frequency $\Omega_z$ {
from the small oscillation of the rms axial size of the BEC upon 
  real time evolution of the mean-field equations (\ref{gp3d}) and (\ref{unibulk}).}
Because of a mixture of frequencies of higher modes,  it was not possible to obtain precisely 
the frequency  $\Omega_\rho$
from a solution of the mean-field equations. The mean-field and the variational results for  $\Omega_z$ are in good agreement 
with  each other. 
These frequencies are practically insensitive 
to a variation of $a_{dd}$ as well as of $\xi$. Hence it may not be very fruitful to study these frequencies 
in the strong-coupling regime in order to extract the parameter $\xi$, specially for a moderate density as in 
this study.

\begin{figure}[!b]
\begin{center}
\includegraphics[width=\linewidth,clip]{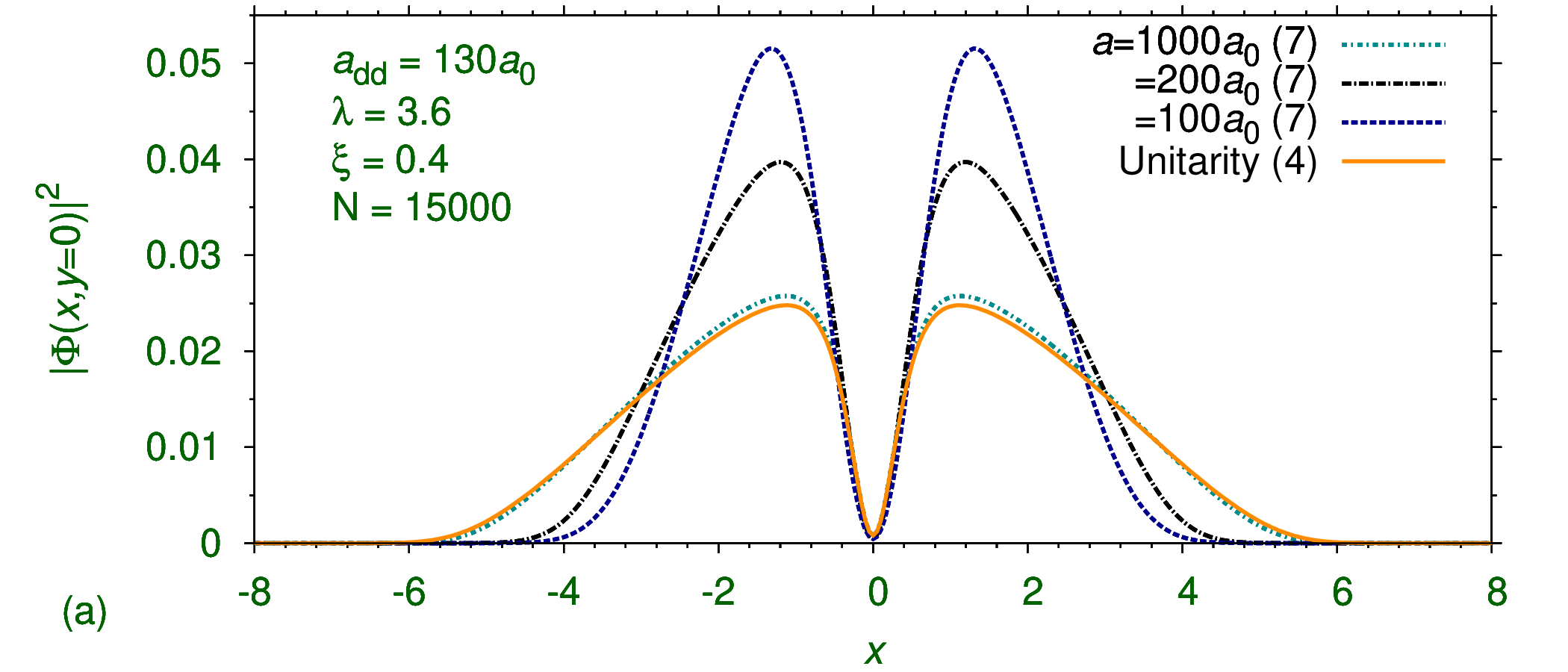}
\includegraphics[width=\linewidth,clip]{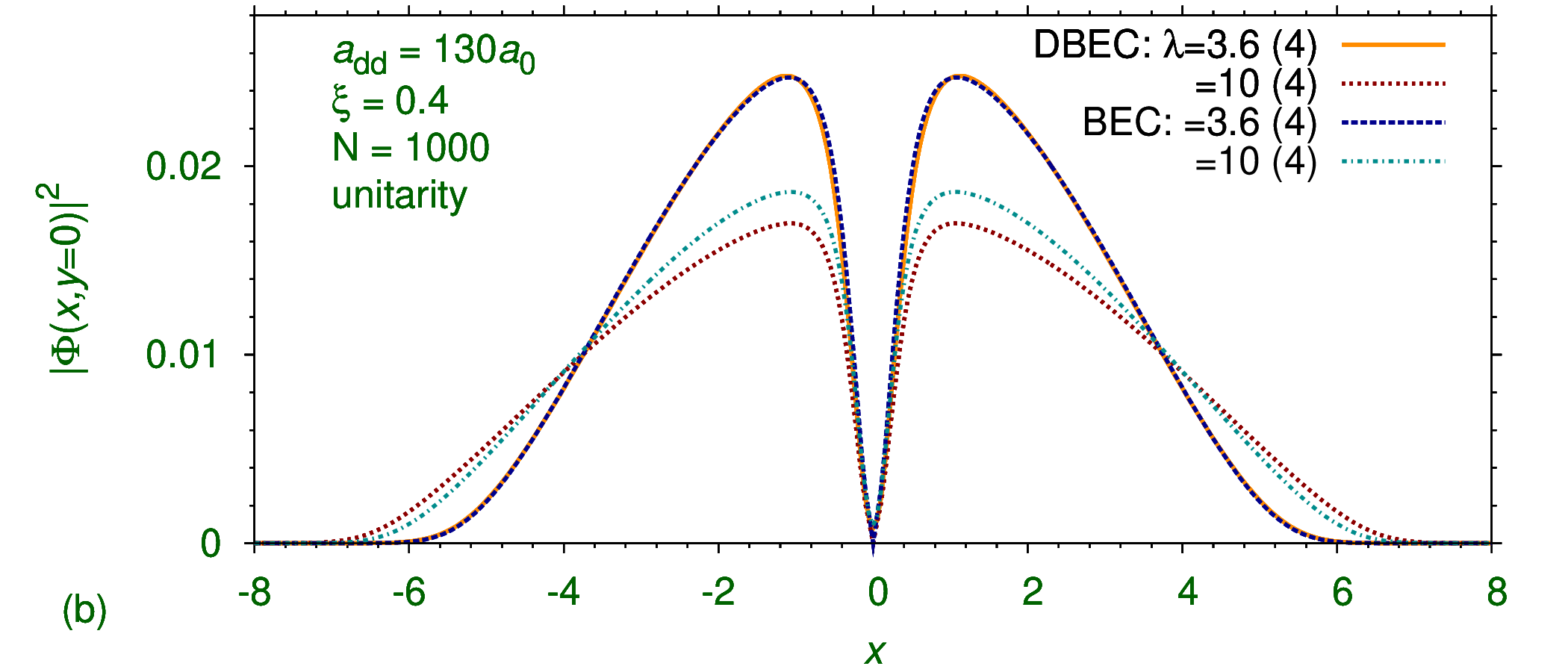}
\end{center}

\caption{(Color online) (a) Radial density along $x$ axis $|\Phi(x,y=0)|^2$ 
of a $^{164}$Dy BEC vortex  of $N=1000$,  $\lambda
=3.6$,  $a_{dd}=130a_0, \nu=1$ and $\xi =0.4$ for  $a=100a_0,
200a_0, 1000a_0$ and at unitarity   using 
the  crossover model (\ref{bulk}), and at unitarity 
(\ref{unibulk}).  
 (b)  The same for  $\xi =0.4, \lambda = 3.6,10$ at unitarity (\ref{unibulk}) for $a_{dd}=130a_0$ (DBEC), and $a_{dd}=0$ (BEC).}

\label{fig4}
\end{figure}

\subsection{Dipolar BEC Vortex}

Next we study the density of a disk-shaped dipolar BEC
vortex of unit angular momentum for strong-coupling 
and demonstrate the sensitivity of the result on the parameter $\xi$.
 In this case the radius of the vortex core is an  observable 
directly related to the healing length \cite{rmp}
of the BEC   and will also be considered. The radius of the vortex 
core $\rho_c$ is defined as the radial distance from the center of the vortex to a point
where the density 
increases to the  maximum value. It is more appropriate to consider the relative radius 
of vortex core defined by $r_0\equiv \rho_c/\langle x \rangle$, which gives the vortex core radius in 
relation to the radial size of the condensate.  It is demonstrated that the relative vortex core radius $r_0$
could be  sensitive to   $\xi$ in the strong-coupling regime and could be useful in deciding 
the value of $\xi$.

In Fig. \ref{fig4} (a), we plot the radial 
density of the BEC vortex along the $x$ axis $|\Phi(x,y=0)|^2$ for $N=1000$, $\lambda =3.6, a_{dd}
= 130a_0,  \xi =0.4, \nu =1$ for  different values of 
scattering length $a$.  
In this figure we show the 
result for  $a=100a_0$ and   at unitarity (\ref{unibulk}) in addition to the results  
for $a=100a_0, 200a_0,$ and $1000a_0$ using the BEC-unitarity 
crossover model (\ref{bulk}).  For small $a$, the 
density obtained using 
 the crossover model (\ref{bulk})
is in agreement with the GP equation (\ref{gpbulk})
 and hence independent of the parameter $\xi$, whereas 
 for large $a$ it 
approximates the unitarity limit (\ref{unibulk}).  In Fig. 
\ref{fig4} (b), 
we show the radial density at unitarity for nondipolar and dipolar BECs
for two values of the trap asymmetry $\lambda = 3.6 $ and 10. The difference between the two densities 
is more pronounced for $\lambda =10$, where the dipolar repulsion is stronger.

\begin{figure}[!t]
\begin{center}
\includegraphics[width=\linewidth,clip]{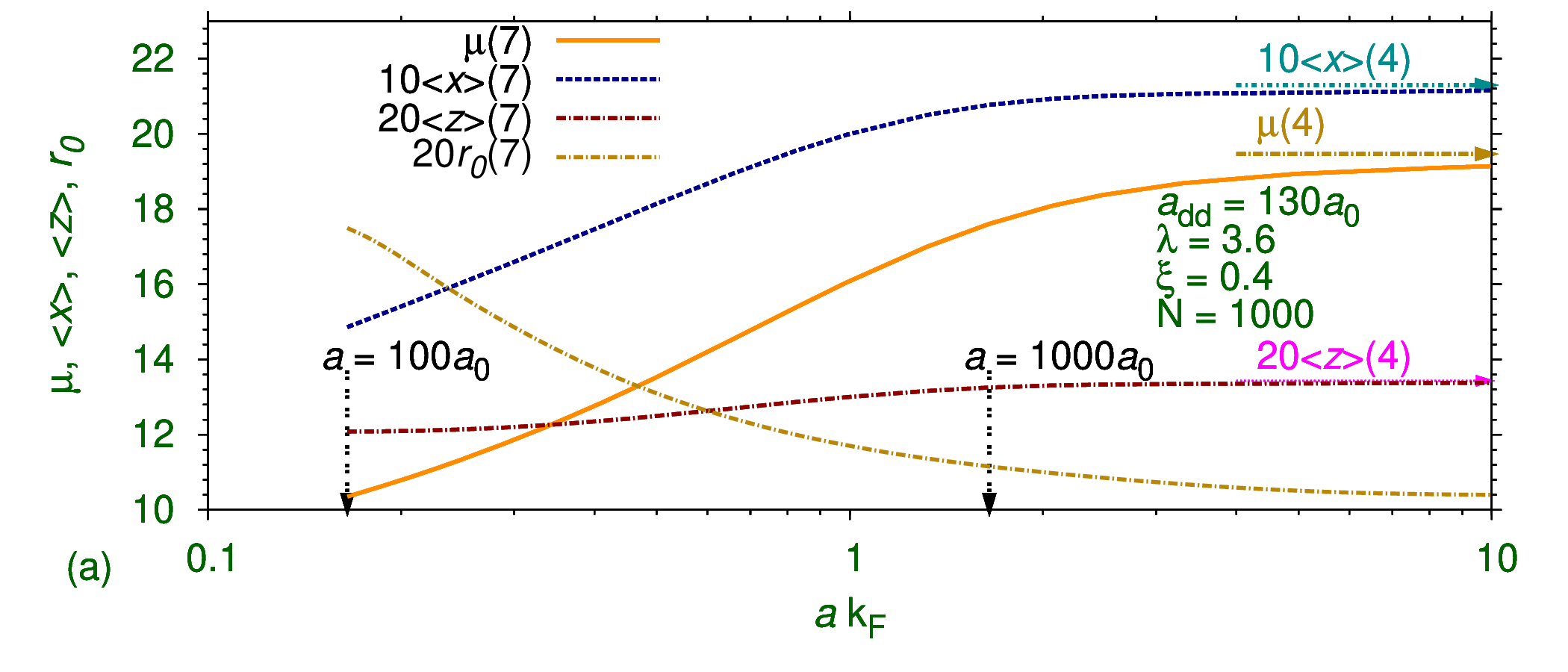}
\includegraphics[width=\linewidth,clip]{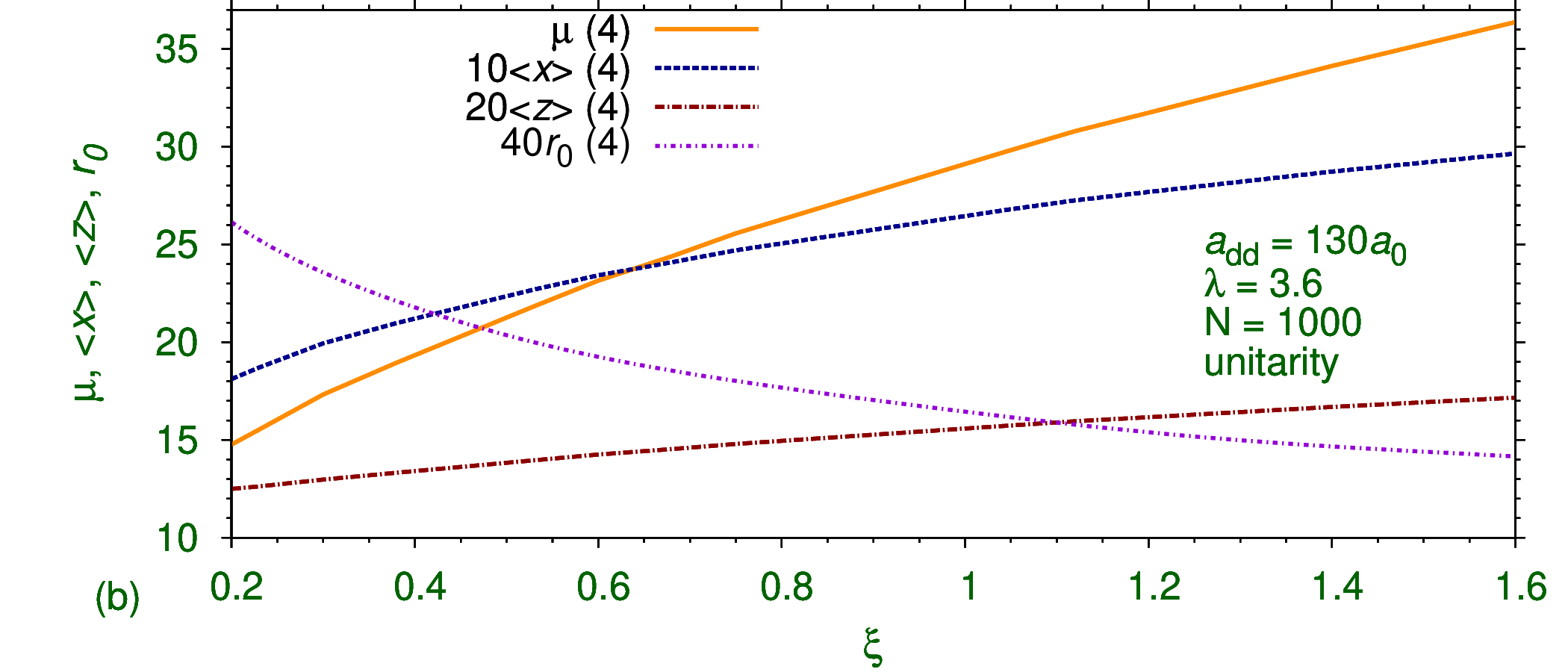}
\end{center}

\caption{(Color online) (a) The   rms sizes $\langle x \rangle$, $\langle z \rangle$,  chemical potential $\mu$, and relative radius
of vortex core $r_0$ 
of the dipolar BEC vortex with $N=1000, a_{dd}
=130a_0, \xi=0.4, \nu=1, \lambda =3.6$ versus 
$ak_F$ using the   crossover model 
(\ref{bulk}),  as well as  at unitarity (\ref{unibulk}) (arrow).
(b) The same  at unitarity   (\ref{unibulk}) versus $\xi$.
 }
\label{fig5}
\end{figure}

In Fig. \ref{fig5} (a) we plot chemical potential $\mu$ and rms sizes $\langle x \rangle$, $\langle z \rangle$
together with the relative radius of vortex core $r_0$ versus $ak_F$ for $N=1000,  a_{dd}=
130a_0, \lambda = 3.6, \xi =0.4, \nu =1$ obtained using the crossover model (\ref{bulk}). 
The result  at   unitarity   (\ref{unibulk}) is also shown. 
{The relative radius of vortex core reduces with the increase of the scattering length. 
Similar reduction of the radius of vortex core was predicted for a nondipolar BEC before \cite{nilsen,skab}.}  
In 
In Fig. \ref{fig5} (b) we plot $\mu$,$\langle x \rangle$, $\langle z \rangle$, $r_0$   versus $\xi$ 
at unitarity  (\ref{unibulk}) for  $N=1000$, $\lambda =3.6, a_{dd}=
130a_0, \lambda = 3.6,  \nu =1$.  As scattering length $a$ increases in Fig. \ref{fig5}  (a) or the parameter $\xi$ increases 
in Fig. \ref{fig5} (b), the system becomes more repulsive leading to a smaller  healing length. Consequently,
the   relative radius of vortex core $r_0$, which is closely related to the healing length,  decreases \cite{rmp}.
The relative radius of vortex core shows much sensitivity to the 
scattering length $a$ and $\xi$. 


\section{Conclusion}


The properties of a BEC at unitarity is controlled by a universal parameter $\xi$ { relating the energies of 
noninteracting and unitary uniform gases}. 
Using the BEC-unitarity crossover model (\ref{bulk}) we studied the properties of a disk-shaped dipolar BEC and dipolar BEC vortex 
in the strong-coupling regime. We find that the density profiles are sensitive to the parameter 
$\xi$ in this regime and  a study of density 
should yield an information about this parameter.
We also studied the frequencies of the fundamental modes of 
radial and axial oscillation of this BEC and find that they are not much sensitive  to $\xi$. 
For a dipolar BEC vortex, in addition 
to density, the relative radius of vortex core is also found to be sensitive to $\xi$ in the strong-coupling regime, 
so that a study of this radius may reveal information about $\xi$.
Also to extract the parameter $\xi$ it is not necessary to study the system
at unitarity. The density profile of the BEC is sensitive   to the parameter 
$\xi$ for  the contact interaction 
lying between the weak-coupling GP and strong-coupling unitarity limits, so that a study in this regime 
should reveal information about this parameter.
In this study we used a dipolar BEC of 15000  $^{164}$Dy atoms in a disk-shaped trap of
anisotropy $\lambda =3.6$, as in the experiment of Ref. \cite{dy2}, and also 
$\lambda =10$. For an experimental study the anisotropy of $\lambda =10$, or larger, and a BEC with
strong dipole   interaction is to be preferred.

\acknowledgments
We thank
FAPESP (Brazil),   CNPq (Brazil),    DST (India),   and CSIR  (India)
for partial support.


\end{document}